\documentclass[prx,aps,twocolumn,superscriptaddress,nofootinbib]{revtex4-1}

\usepackage{amsmath,graphicx,epsfig}
\usepackage{epstopdf}
\usepackage{euscript}
\usepackage{amsfonts}
\usepackage{amssymb}
\usepackage{float}

\def\prn#1{{\left(#1\right)}}

\def\sbrk#1{{\left[#1\right]}}

\def\bra#1{{\langle#1|}}

\def\cg(#1,#2)(#3,#4)(#5,#6){\bra{#1,#2,#3,#4}#5,#6\rangle}

\def\ts#1{{_{\mbox{\scriptsize #1}}}}

\def\threej(#1,#2)(#3,#4)(#5,#6){\begin{pmatrix}#1&#3&#5\\#2&#4&#6\end{pmatrix}}
\def\sixj(#1,#2,#3)(#4,#5,#6){\begin{Bmatrix}#1&#2&#3\\#4&#5&#6\end{Bmatrix}}
\def\ninej(#1,#2,#3)(#4,#5,#6)(#7,#8,#9){\begin{Bmatrix}#1&#2&#3\\#4&#5&#6\\#7&#8&#9\end{Bmatrix}}

\def\mc{\mathcal}

\newlength{\defbaselineskip}
\newcommand{\setlinespacing}[1]%
           {\setlength{\baselineskip}{#1 \defbaselineskip}}

\begin{document}

\title{Overview of the Cosmic Axion Spin Precession Experiment (CASPEr)}

\author{D.~F.~Jackson Kimball}
\email{derek.jacksonkimball@csueastbay.edu}
\affiliation{California State University - East Bay, Hayward, California, USA}

\author{S.~Afach}
\affiliation{Helmholtz Institute, Mainz, Germany}

\author{D.~Aybas}
\affiliation{Boston University, Boston, Massachusetts, USA}

\author{J.~W.~Blanchard}
\affiliation{Helmholtz Institute, Mainz, Germany}

\author{D.~Budker}
\affiliation{Johannes Gutenberg-Universit\"at, Mainz, Germany}
\affiliation{Helmholtz Institute, Mainz, Germany}
\affiliation{University of California, Berkeley, California, USA}
\affiliation{Lawrence Berkeley National Laboratory, Berkeley, California, USA}

\author{G.~Centers}
\affiliation{Johannes Gutenberg-Universit\"at, Mainz, Germany}
\affiliation{Helmholtz Institute, Mainz, Germany}

\author{M.~Engler}
\affiliation{Johannes Gutenberg-Universit\"at, Mainz, Germany}
\affiliation{Helmholtz Institute, Mainz, Germany}

\author{N.~L.~Figueroa}
\affiliation{Johannes Gutenberg-Universit\"at, Mainz, Germany}
\affiliation{Helmholtz Institute, Mainz, Germany}

\author{A.~Garcon}
\affiliation{Johannes Gutenberg-Universit\"at, Mainz, Germany}
\affiliation{Helmholtz Institute, Mainz, Germany}

\author{P.~W.~Graham}
\affiliation{Stanford University, Stanford, California, USA}

\author{H.~Luo}
\affiliation{Shanghai Institute of Ceramics, Chinese Academy of Sciences, China}

\author{S.~Rajendran}
\affiliation{University of California, Berkeley, California, USA}

\author{M.~G.~Sendra}
\affiliation{Johannes Gutenberg-Universit\"at, Mainz, Germany}
\affiliation{Helmholtz Institute, Mainz, Germany}

\author{A.~O.~Sushkov}
\affiliation{Boston University, Boston, Massachusetts, USA}

\author{T.~Wang}
\affiliation{University of California, Berkeley, California, USA}

\author{A.~Wickenbrock}
\affiliation{Johannes Gutenberg-Universit\"at, Mainz, Germany}
\affiliation{Helmholtz Institute, Mainz, Germany}

\author{A.~Wilzewski}
\affiliation{Johannes Gutenberg-Universit\"at, Mainz, Germany}
\affiliation{Helmholtz Institute, Mainz, Germany}

\author{T.~Wu}
\affiliation{Helmholtz Institute, Mainz, Germany}

\date{\today}

\begin{abstract}
An overview of our experimental program to search for axion and axion-like-particle (ALP) dark matter using nuclear magnetic resonance (NMR) techniques is presented. An oscillating axion field can exert a time-varying torque on nuclear spins either directly or via generation of an oscillating nuclear electric dipole moment (EDM). Magnetic resonance techniques can be used to detect such an effect. The first-generation experiments explore many decades of ALP parameter space beyond the current astrophysical and laboratory bounds. It is anticipated that future versions of the experiments will be sensitive to the axions associated with quantum chromodynamics (QCD) having masses $\lesssim 10^{-9}~{\rm eV}/c^2$.
\end{abstract}

\maketitle

\section{Coupling of dark-matter axions and axion-like-particles (ALPs) to spins}
\label{sec:spin-coupling}

The original motivation for introducing axions (pseudoscalar particles) emerged from an elegant solution to the strong-$CP$ problem\footnote{$CP$ refers to the combined symmetry with respect to charge-conjugation ($C$), transformation between matter and anti-matter, and spatial inversion, i.e., parity transformation ($P$).} by Peccei and Quinn \cite{Pec77a,Pec77b}. The strong-$CP$ problem is the observation that the theory of quantum chromodynamics (QCD) requires extreme fine-tuning of parameters in order to be reconciled with the observation that the strong interaction is found experimentally to respect $CP$ symmetry to a high degree, as evidenced by experimental constraints on the neutron permanent electric dipole moment (EDM) \cite{Bak06}. The so-called ``QCD axion'' emerges when the strong-$CP$ problem is resolved by introducing a new symmetry that is broken at a high energy scale $f_a$, possibly as high as the Planck scale \cite{Wei78,Wil78}. Since the original proposal of Peccei and Quinn \cite{Pec77a,Pec77b}, the axion concept has had a broad and significant impact on theoretical physics. Axion-like-particles (ALPs) emerge naturally whenever a global symmetry is broken \cite{Wei78,Wil78,Kim79,Shi80,Din81,Zhi80}: such global symmetry breaking is a ubiquitous feature of beyond-Standard-Model physics, including grand unified theories (GUTs), models with extra dimensions, and string theory \cite{Svr06,Arv10}. In the following, we use the term axion to refer to both the QCD axion and ALPs as long as the discussion pertains to both and make distinctions when necessary.

Interactions of the axion field $a$, via QCD or other mechanisms in the case of ALPs \cite{Din83,Pre83}, generate a potential energy density $\sim m_a^2 c^2 a^2/(2\hbar^2)$, where $m_a$ is the axion mass. Initial displacement of the axion field from the minimum of this potential results in oscillations of the axion field at the Compton frequency
\begin{equation}
\omega_a = \frac{m_a c^2}{\hbar}~.
\label{Eq:Compton-frequency}
\end{equation}
The energy density in these oscillations can constitute the mass-energy associated with dark matter. The temporal coherence of oscillations of the dark-matter axion field $a(\vec{r},t)$ observed in a terrestrial experiment is limited by relative motion through random spatial fluctuations of the field. The size of such fluctuations corresponds to the axion de Broglie wavelength $\lambda\ts{dB}$, thus the coherence time is
\begin{equation}
\tau_a \approx \frac{\lambda\ts{dB}}{v} \approx \frac{2\pi\hbar}{m_a v^2}~,
\label{Eq:axion-coherence-time}
\end{equation}
where $v \sim 10^{-3}c$ is the galactic virial velocity of the dark-matter axion field. Therefore the ``quality factor'' $Q$ corresponding to a dark-matter axion field is given by
\begin{equation}
Q = \frac{\omega_a\tau_a}{2\pi} \approx \prn{\frac{c}{v}}^2 \approx 10^6~.
\label{Eq:Q}
\end{equation}
The $Q$ of the axion field can also be understood by noting that due to the second-order Doppler shift that arises from the contribution of the kinetic energy of the axions to their total energy, axions moving at $v$ relative to a detector are measured to have a frequency
\begin{equation}
\omega' \approx \omega_a \prn{ 1 + \frac{v^2}{2c^2} }~.
\label{Eq:Doppler-shifted-frequency}
\end{equation}
Thus the spread in axion velocities leads to a spread in observed frequencies, giving the axion field the $Q$ shown in Eq.~(\ref{Eq:Q}).

The axion field seen by a detector on the Earth is $a(t) = a_0 \cos(\omega' t)$, a field oscillating at $\approx \omega_a$, and the amplitude of the field $a_0$ can be estimated by assuming the energy of the axion field comprises the totality of the local dark matter energy density $\rho\ts{DM} \approx 0.4~{\rm GeV/cm^3}$ \cite{Ber98,Jun96,Sof01}:
\begin{equation}
\rho\ts{DM} \approx \frac{c^2}{2\hbar^2} m_a^2a_0^2~.
\label{Eq:a0-from-DM-density}
\end{equation}

To detect such an oscillating axion field with a terrestrial sensor, one searches for (non-gravitational) interactions of $a(\vec{r},t)$ with Standard Model fields and particles. Axion/ALP fields $a(\vec{r},t)$ possess three such interactions, in general, that can be described by the Lagrangians (given in natural units where $\hbar=c=1$) \cite{Gra13}
\begin{align}
\mc{L}\ts{EM} &\approx g_{a\gamma\gamma} a(\vec{r},t) F_{\mu\nu} \tilde{F}^{\mu\nu}~,  \label{Eq:axion-EM-coupling} \\
\mc{L}\ts{EDM} &\approx -\frac{i}{2} g_{d} a(\vec{r},t) \overline{\Psi}_n \sigma_{\mu\nu} \gamma_5 \Psi_n F^{\mu\nu}~, \label{Eq:axion-EDM-coupling} \\
\mc{L}\ts{spin} &\approx g_{aNN} \sbrk{ \partial_\mu a(\vec{r},t)  } \overline{\Psi}_n \gamma^{\mu} \gamma_5 \Psi_n~, \label{Eq:axion-spin-coupling}
\end{align}
where $g_{a\gamma\gamma}$ parameterizes the axion-photon coupling, $g_{d}$ parameterizes the axion-gluon coupling that generates nuclear EDMs, $g_{aNN}$ parameterizes the coupling to nuclear spins,\footnote{There can be a similar coupling to electron spins.} $F_{\mu\nu}$ is the electromagnetic field tensor, $\Psi_n$ is the nucleon wave function, and $\sigma$ and $\gamma$ are the standard Dirac matrices. Note that the coupling constants $g_{a\gamma\gamma}$, $g_{d}$, and $g_{aNN}$ are proportional to $1/f_a$ \cite{Gra13}. The axion-photon interaction described by Eq.~(\ref{Eq:axion-EM-coupling}) is used in a variety of ``haloscope'' experiments\footnote{Haloscope experiments {\emph{directly}} detect the dark matter from the galactic halo \cite{Sik83,Sik85}. Complementary approaches include (1) ``helioscope'' experiments that search for axions emitted by the Sun; (2) ``light-shining-through-walls'' experiments where axions are created from an intense laser light field passing through a strong magnetic field (which facilitates mixing between photons and axions) and then detected by converting them back to photons after they cross a wall that is transparent to them but completely blocks the light; and (3) {\emph{indirect}} experiments that search for modifications of known interactions due to exchange of virtual axions. Constraints on axions and ALPs can also be obtained from astrophysical observations, see for example Refs.~\cite{Vys78,Ber91,Sak96,Khl99,Raf88,Raf95,Raf12,Blu16,Cha18}. These alternative approaches are reviewed in Refs.~\cite{Saf18,Gra15,Raf99}.} to search for axion dark matter as discussed elsewhere in this volume, such as the Axion Dark Matter eXperiment (ADMX) \cite{Asz01,Asz10}, the Haloscope At Yale Sensitive To Axion Cold dark matter (HAYSTAC) \cite{Bru17}, and CAPP's (Center for Axion and Precision Physics Research) Ultra Low Temperature Axion Search in Korea (CULTASK) \cite{You16}. In contrast to other haloscope experiments searching for axion-photon interactions, the Cosmic Axion Spin Precession Experiment (CASPEr, see Ref.~\cite{Bud14}) exploits the axion couplings to nuclear spins described by Eqs.~(\ref{Eq:axion-EDM-coupling}) and (\ref{Eq:axion-spin-coupling}).

The Lagrangian $\mc{L}\ts{EDM}$ describes an oscillating nuclear EDM $\vec{d}_n(t)$, generated by $a(t)$ along the direction of the nuclear spin $\hat{\vec{\sigma}}_n$, given in natural units by the expression
\begin{equation}
\vec{d}_n(t) = g_d a_0 \cos(\omega_a t)\hat{\vec{\sigma}}_n~,
\label{Eq:oscillating-EDM}
\end{equation}
that interacts with an external electric field $\vec{E}$. The oscillating EDM amplitude can be estimated from Eq.~(\ref{Eq:a0-from-DM-density}):
\begin{align}
d_n &= g_d a_0 \approx \frac{g_d}{m_a} \sqrt{ 2\rho\ts{DM} }~, \label{Eq:EDM-amplitude-1}\\
&\approx 6 \times 10^{-25}~{\rm e \cdot cm} \times \frac{g_d\sbrk{ {\rm GeV^{-2} } }}{m_a\sbrk{ {\rm eV } }}~, \label{Eq:EDM-amplitude-2}
\end{align}
where $\sbrk{\cdots}$ indicates the units of the respective quantity. The non-relativistic Hamiltonian describing this interaction is
\begin{equation}
H\ts{EDM} = -\vec{d}_n(t) \cdot \vec{E}~,
\label{Eq:EDM-Hamiltonian}
\end{equation}
and there is a corresponding spin-torque $\vec{\tau}\ts{EDM}$
\begin{equation}
\vec{\tau}\ts{EDM} = \vec{d}_n(t) \times \vec{E}~.
\label{Eq:EDM-torque}
\end{equation}

The Lagrangian $\mc{L}\ts{spin}$ results in a non-relativistic Hamiltonian (in natural units)
\begin{equation}
H\ts{spin} = g_{aNN} \vec{\nabla} a(\vec{r},t) \cdot \hat{\vec{\sigma}}_n~,
\label{Eq:axion-spin-Hamiltonian}
\end{equation}
which describes the interaction of nuclear spins with an oscillating ``pseudo-magnetic field'' generated by the gradient of the axion field.  The magnitude of the gradient $\vec{\nabla} a(\vec{r},t)$ can be estimated by noting that since $\vec{p} = -i\hbar \vec{\nabla}$,
\begin{equation}
\left|\vec{\nabla} a \right| \approx \frac{m_a v}{\hbar} a_0~.
\label{Eq:axion-gradient}
\end{equation}
This is the so-called ``axion wind'' which acts as a pseudo-magnetic field directed along $\vec{v}$ \cite{Gra13}.  The resultant Hamiltonian is
\begin{align}
H\ts{wind} &\approx g_{aNN} m_a a_0 \cos(\omega_a t) \vec{v} \cdot \hat{\vec{\sigma}}_n~, \label{Eq:wind-Hamiltonian-1}\\
&\approx g_{aNN} \sqrt{2 \hbar^3 c \rho\ts{DM} } \cos(\omega_a t) \vec{v} \cdot \hat{\vec{\sigma}}_n~,  \label{Eq:wind-Hamiltonian-2}
\end{align}
where the amplitude of the axion field assumed in Eq.~(\ref{Eq:a0-from-DM-density}) was used in Eq.~(\ref{Eq:wind-Hamiltonian-2}), in which $H\ts{wind}$ is expressed in Gaussian units. Given a nucleus with a particular gyromagnetic ratio $\gamma_n = g_n \mu_N/\hbar$, where $g_n$ is the nuclear Land\'e factor and $\mu_N$ is the nuclear magneton, the amplitude $B_a$ of the oscillating pseudo-magnetic field produced by the axion field can be estimated to be
\begin{align}
B_a &\approx 10^{-3} \times \frac{g_{aNN}}{\hbar\gamma_n} \sqrt{ 2\hbar^3c^3 \rho\ts{DM} }~, \label{Eq:pseudo-B-1} \\
B_a\sbrk{\rm T } &\approx 10^{-7} \times \frac{g_{aNN}\sbrk{\rm GeV^{-1} }}{g_n}~, \label{Eq:pseudo-B-2}
\end{align}
where we have assumed that $v \approx 10^{-3}c$, the virial velocity of the axions. Analogous to the case of the axion-induced oscillating nuclear EDM $\vec{d}_n(t)$ discussed above, here we have an oscillating pseudo-magnetic field $\vec{B}_a(t)$ that interacts with the nuclear magnetic dipole moment $\vec{\mu}_n$:
\begin{equation}
H\ts{wind} = -\vec{\mu}_n \cdot \vec{B}_a(t)~,
\label{Eq:wind-Hamiltonian-3}
\end{equation}
and a corresponding spin-torque $\vec{\tau}\ts{wind}$
\begin{equation}
\vec{\tau}\ts{wind} = \vec{\mu}_n \times \vec{B}_a(t)~.
\label{Eq:wind-torque}
\end{equation}

On the one hand, the interactions described by Eqs.~(\ref{Eq:EDM-Hamiltonian},\ref{Eq:EDM-torque}) and (\ref{Eq:wind-Hamiltonian-3},\ref{Eq:wind-torque}) have similar signatures and thus suggest similar experimental approaches: in both cases the goal of an axion-dark-matter experiment would be to detect an oscillating nuclear-spin-dependent energy shift (or, analogously, an oscillating torque on nuclear spins). The well-developed techniques of nuclear magnetic resonance (NMR) are ideally suited to this task. On the other hand, there is an essential difference between the signatures of $\mc{L}\ts{EDM}$ and $\mc{L}\ts{spin}$, namely in the case of $\mc{L}\ts{EDM}$ an electric field $\vec{E}$ is required for observation of the effect. For this reason, the CASPEr experimental program is divided into two branches: CASPEr Electric, which searches for an oscillating EDM $\vec{d}_n(t)$, and CASPEr Wind, which searches for an oscillating pseudo-magnetic field $\vec{B}_a(t)$ \cite{Bud14}.

\section{General Principles of the CASPEr Nuclear Magnetic Resonance (NMR) Approach}
\label{sec:NMR}

\begin{figure}
\includegraphics[scale=.45]{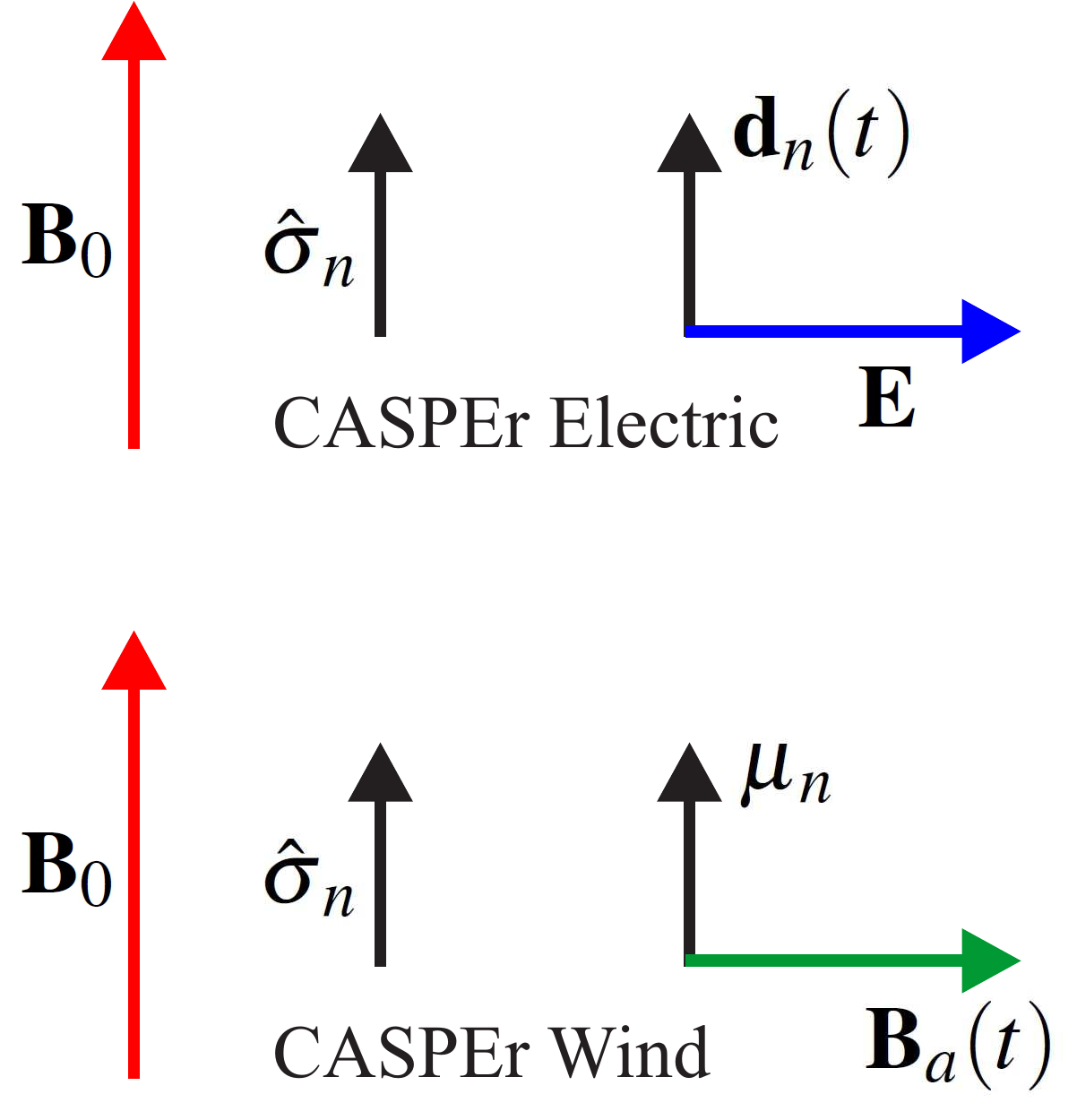}
\caption{Experimental geometries for CASPEr Electric (top) and CASPEr Wind (bottom). In both cases, the nuclear spins $\hat{\vec{\sigma}}_n$ are oriented along a leading magnetic field $\vec{B}_0$. An oscillating torque, $\vec{\tau}\ts{EDM} = \vec{d}_n(t) \times \vec{E}$ in the case of CASPEr Electric and $\vec{\tau}\ts{wind} = \vec{\mu}_n \times \vec{B}_a(t)$ in the case of CASPEr Wind, tips the nuclear spins away from $\vec{B}_0$ if the Larmor frequency $\Omega_L$ matches $\omega_a$.}
\label{Fig:ExptGeometry}
\end{figure}

\begin{figure}
\includegraphics[scale=.45]{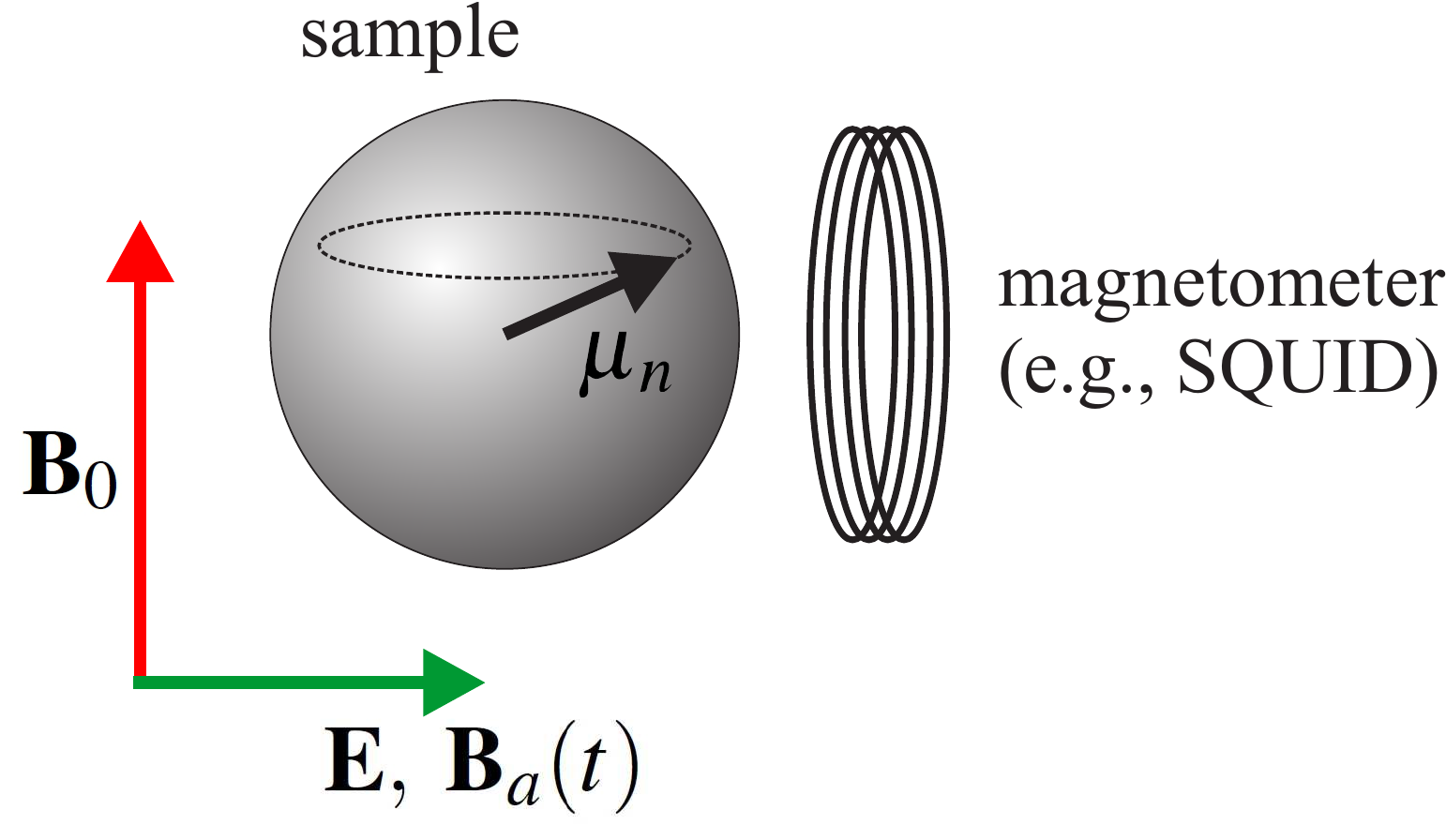}
\caption{Schematic diagram of the CASPEr experiment. When $\Omega_L \approx \omega_a$, the nuclear spins in the sample are tipped away from their initial orientation along $\vec{B}_0$ due to the axion-induced torque. The precessing magnetization at $\Omega_L$ can be detected with a magnetometer such as a SQUID placed near the sample.}
\label{Fig:ExptSchematic}
\end{figure}

Both CASPEr Electric and CASPEr Wind use NMR techniques to search for axion dark matter. The experimental geometries are shown in Fig.~\ref{Fig:ExptGeometry}. The leading field $\vec{B}_0$ determines the Larmor frequency,
\begin{equation}
\Omega_L = \gamma_n B_0~,
\end{equation}
for the nuclear spins, which are initially oriented along $\vec{B}_0$. If $\Omega_L \neq \omega_a$, then the time-dependent torques given by Eqs.~(\ref{Eq:EDM-torque}) and (\ref{Eq:wind-torque}) average out and there is no appreciable effect on the spins. However, when $\Omega_L \approx \omega_a$, a resonance occurs and the spins are tilted away from the direction of $\vec{B}_0$. Viewed from the frame rotating with the spins at $\Omega_L$, the effective leading field in the rotating frame goes to zero,
\begin{equation}
\vec{B}\ts{eff}\prn{ \rm rot } = \vec{B}_0 - \frac{\boldmath{\Omega}_L}{\gamma_n} = 0~.
\label{Eq:field-in-rotating-frame}
\end{equation}
The axion-induced torque oscillating in the laboratory frame has a static component in the rotating frame when $\Omega_L \approx \omega_a$, and thus is able to tilt $\hat{\vec{\sigma}}_n$ away from the direction of $\vec{B}_0$ (see, for example, Problem 2.6 of Ref.~\cite{Bud08} for a tutorial discussion of this well-known effect). In the laboratory frame, the tilted nuclear spins are observed to precess in $\vec{B}_0$. This axion-induced nuclear spin precession at $\Omega_L \approx \omega_a$ is the signature of the axion dark matter detected in both CASPEr Electric and CASPEr Wind. The precessing magnetization can be measured, for example, by induction through a pick-up loop or with a Superconducting QUantum Interference Device (SQUID), see Fig.~\ref{Fig:ExptSchematic}.

The amplitude of the NMR signal is proportional to the tilt angle $\varphi$ of the spins (since under our conditions $\varphi \ll 1$), which for CASPEr Electric is given by
\begin{equation}
\varphi\ts{EDM} \approx \frac{d_n E T_2}{\hbar} \approx g_d \frac{E T_2}{\hbar m_a} \sqrt{ \frac{2\hbar^3}{c} \rho\ts{DM} }~,
\label{Eq:tilt-angle-EDM}
\end{equation}
and for CASPEr Wind is given by
\begin{equation}
\varphi\ts{wind} \approx \frac{\mu_n B_a T_2}{\hbar} \approx g_{aNN} \frac{\mu_n T_2}{1000 \hbar^2 \gamma_n} \sqrt{ 2\hbar^3c^3 \rho\ts{DM} }~,
\label{Eq:tilt-angle-wind}
\end{equation}
where $T_2$ is the spin-precession (transverse) coherence time. The ultimate limit on $T_2$ is the axion coherence time $\tau_a$ [Eq.~(\ref{Eq:axion-coherence-time})], but in many cases $T_2$ is determined by the material properties of the nuclear spin sample.

A key to CASPEr's sensitivity is the coherent ``amplification'' of the effects of the axion dark matter field through a large number of polarized nuclear spins. For example, in the case of CASPEr Wind, the spins tip by an angle $\varphi\ts{wind}$ during the coherence time [Eq.~(\ref{Eq:tilt-angle-wind})]. Consider a spherical sample with a polarized nuclear spin density of $n$. The amplitude of the oscillating field $B_1$ generated by the precessing magnetization that is detected by the magnetometer as shown in Fig.~\ref{Fig:ExptSchematic} is
\begin{equation}
B_1 \approx \prn{ \frac{8\pi}{3} \mu_n n } \varphi\ts{wind} \approx \prn{ \frac{8\pi}{3\hbar} \mu_n^2 n T_2 } B_a = \eta B_a~,
\end{equation}
where the factor $\eta$ is the effective enhancement of the measured oscillating field $B_1$ over the pseudo-magnetic field $B_a$ generated by the axion dark matter field. The enhancement factor $\eta$ can be considerable: for example, a fully polarized sample of liquid $^{129}$Xe has $n \approx 10^{22}~{\rm spins/cm^3}$ and can have $T_2 \gtrsim 1000~{\rm s}$ \cite{Rom01}, which gives $\eta \gtrsim 10^6$.

The axion mass $m_a$ to which an NMR experiment of this type is sensitive is set by the resonance frequency $\Omega_L=\omega_a$ (although note that Ref.~\cite{Gar17} discusses an alternative broadband method suitable for very low $m_a$). Thus there is an upper limit to the possible range of axion masses explored by CASPEr, given by the largest dc magnetic field achievable in laboratories ($\sim 30~{\rm T}$) which corresponds to $\omega_a \sim 400~{\rm MHz}$ and $m_ac^2 \sim 10^{-6}~{\rm eV}$. The procedure to search for axions of different masses (with an exception for very small $\omega_a$ as discussed in Ref.~\cite{Gar17}) is to scan $B_0$ in increments proportional to the width of the NMR resonance ($\approx 1/T_2$). The integration time at each point in the magnetic field scan should be $\gtrsim T_2$ in order to take full advantage of the coherent build-up of tilted magnetization.

\section{Low mass ($\lesssim 1~{\rm \mu eV}$) axions and cosmology}
\label{sec:LowMassAxions}

As noted above, CASPEr searches for axions with masses $m_ac^2 \lesssim 10^{-6}~{\rm eV}$. It is worth mentioning that cosmological constraints have been considered for QCD axions with masses below $10^{-6}~{\rm eV}$ \cite{Din83,Pre83}. However, these constraints are highly dependent upon assumptions about unknown initial conditions of the universe. Such lighter mass QCD axions are not ruled out either by experimental or astrophysical observations. Cosmological arguments suggesting that $m_ac^2 \gtrsim 10^{-6}~{\rm eV}$ are based on a particular scenario for the earliest epochs in the universe, a time about which we know little.

The original argument posited that QCD axions with masses below $10^{-6}~{\rm eV}$ produced too much dark matter and thereby ``over-closed'' the universe \cite{Din83,Pre83}. In the absence of additional physics, this statement is true if inflation precedes axion production. On the other hand, if inflation occurs after axion production, the axion abundance in the universe depends sensitively upon the unknown initial ``misalignment'' angle of the axion field \cite{Lin88}, the displacement of the axion field from the local minimum of its potential. If the misalignment angle is sufficiently small overclosure can be avoided. An initial misalignment angle $\ll 1$ can be naturally explained by anthropic considerations: namely the idea that there is an initial random distribution of misalignment angles in the universe so that, of course, humans exist in regions of the universe where the misalignment angle is such that life is possible \cite{Lin88,Lin07}. Note that such anthropic arguments are a generic explanation for the observed coincidence in the energy density of dark matter and the baryon content of the universe \cite{McD13}. This coincidence is particularly suggestive for axion dark matter since the axion and the baryon abundances arise from completely different physics. It should also be noted that such ``tuned'' values of the axion misalignment angle also naturally emerge in models such as the relaxion scenario where the axion solves the hierarchy problem \cite{Gra15relaxion}.

It is also crucial to note that the possible cosmological constraints discussed above do not apply to ALPs. In summary, not only do axions with masses $\lesssim 10^{-6}~{\rm eV}$ fit well within the landscape of theoretical particle physics, but they also probe particularly interesting physics as they arise from symmetry breaking at the GUT and Planck scales \cite{Svr06,Arv10,Gra13,Saf18,Gra15,Bud14}.

\section{CASPEr Electric}
\label{sec:CASPEr-Electric}

The sample to be employed in the CASPEr Electric experiment is a ferroelectric crystal such as those considered for static nuclear EDM experiments \cite{Sha68,Leg78,Muk05,Bud06}; recently experiments were carried out with paramagnetic ferroelectrics to search for the electron EDM \cite{Sus10,Rus10,Eck12}. A ferroelectric crystal possesses permanent electric polarization that creates a strong effective internal electric field $\vec{E}^*$ (on the order of 100~MV/cm)\footnote{The mechanism generating $\vec{E}^*$ is similar to that generating the effective electric fields experienced by EDMs in polar molecules \cite{Bar14}.} with which an axion-induced nuclear EDM can interact. Searching for an axion-induced EDM requires a heavy atom in order to minimize the Schiff screening, which arises because the positions of charged particles in the sample tend to adjust themselves to cancel internal electric fields \cite{Sch63,San65}. In ferroelectric crystals, the effective energy shift produced by a nuclear EDM is given by $\epsilon_S d_n E^*$, where $\epsilon_S$ is the Schiff suppression factor \cite{Sha68,Leg78,Muk05}.

It is advantageous for NMR detection that the heavy atom has a large nuclear magnetic moment, and for minimizing spin relaxation it is desirable that the nucleus has spin $I = 1/2$ \cite{Sha68,Leg78,Muk05}. The sample material must also have non-centrosymmetric crystal structure in order to provide the effective electric field $\vec{E}^*$. Based on these requirements, we have identified several potential samples to investigate: lead germanate (PGO, ${\rm Pb_5Ge_3O_{11} }$), lead titanate (PT, $\rm{ PbTiO_3 }$), lead zirconate titanate (PZT, $\rm{ PbZr_yTi_{1-y}O_3 }$),  lanthanum-doped lead zirconium titanate (PLZT, $\rm{ Pb_xLa_{1-x}Zr_yTi_{1-y}O_3 }$), lead magnesium niobate-lead titanate [PMN-PT, $(1-x){\rm PbMg_{1/3}Nb_{2/3}O_3} – (x){\rm PbTiO_3}$], cadmium titanate ($\rm{ CdTiO_3 }$), and paraelectric materials such as ${\rm SrTiO_3}$, ${\rm LiTaO_3}$, and ${\rm KTaO_3}$. The different samples have different potential advantages, such as tunable ferroelectric properties and optical transparency (which may enable optical control of nuclear spin relaxation or even optical pumping of nuclear spins \cite{Mol68,Hay08}).

\begin{figure*}
\includegraphics[scale=.45]{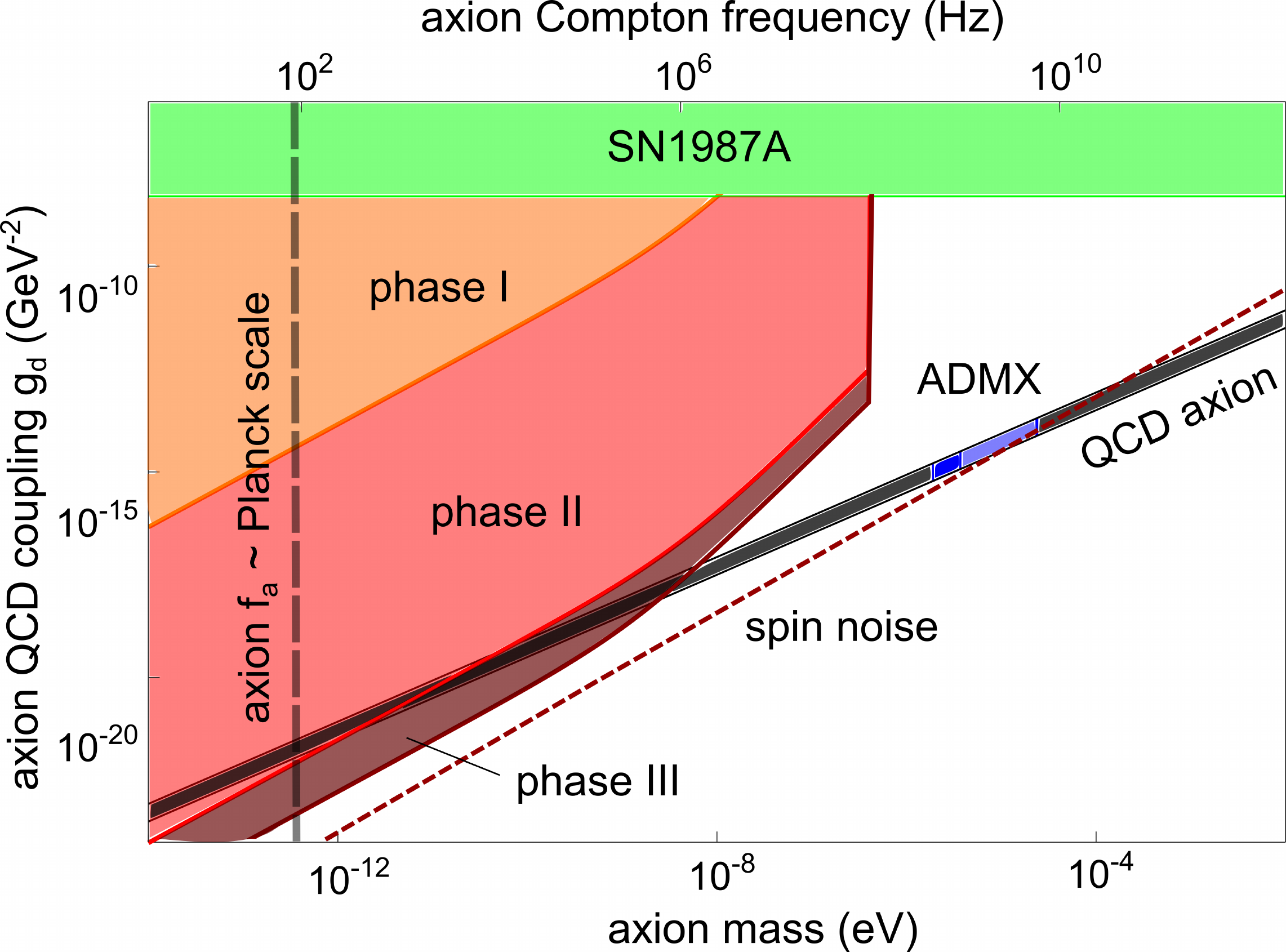}
\caption{Experimental reach of CASPEr Electric. The green band is excluded by astrophysical observations \cite{Gra15,Raf99}. The blue region shows the axion mass range covered by ADMX and HAYSTAC. Orange, red, and maroon regions show sensitivity projections explained in text. Phases II and III reach the QCD axion coupling strength. The fundamental quantum sensitivity limit is given by magnetization noise, shown by the dashed red line. The vertical dashed gray line indicates the mass $m_a$ and frequency $\omega_a$ corresponding to axions generated by symmetry breaking at the Planck scale. See Ref.~\cite{Bud14} for details of these estimates.}
\label{Fig:CASPEr-Electric-sensitivity}
\end{figure*}

Phase I of CASPEr Electric will use thermally polarized $^{207}$Pb nuclear spins ($I = 1/2$, 22\% abundance) in ferroelectric PMN-PT crystals at cryogenic temperatures. The ferroelectric displacement of the Pb ion with respect to its surrounding oxygen cage gives rise to a large effective electric field ${E}^* \approx 3 \times 10^8~{\rm V/cm}$ \cite{Muk05}. The transverse coherence time for the $^{207}$Pb nuclear spins in single-crystal PMN-PT samples is expected to be $T_2 \approx 10^{-3}~{\rm s}$, similar to that observed for $\rm{ PbTiO_3 }$ \cite{Bou08}. Thus over the entire range of $\omega_a$ values to be searched, $T_2 \ll \tau_a$. The longitudinal nuclear spin polarization time is expected to be $T_1 \approx 1000~{\rm s}$ at cryogenic temperatures \cite{Bou08}, and so the strategy is to start at a relatively high magnetic field $B_0 \sim 10~{\rm T}$ and then ramp down $B_0$ over a time scale $\sim T_1$ while there remains significant thermal spin polarization. Note that in this scheme there is a lower bound on $B_0$ due to relaxation caused by paramagnetic impurities in the PMN-PT sample, i.e., when $B_0$ becomes too small local magnetic fields due to paramagnetic impurities cause spins to dephase with respect to one another.

The amplitude of the transverse precessing component of magnetization $M$ that arises due to interaction with the dark matter axion field, for a spin density $n$ and polarization $P$, is given by
\begin{align}
    M &\approx n P \mu_n \varphi\ts{EDM} \approx  n P \mu_n \frac{\epsilon_S d_n E^* T_2}{\hbar} \label{Eq:magnetization-EDM-2} \\
      &\approx n P \mu_n \frac{\epsilon_S E^* T_2}{\hbar} \frac{g_d}{m_a} \sqrt{ \frac{2\hbar^3}{c} \rho\ts{DM} }~, \label{Eq:magnetization-EDM-3}
\end{align}
where Eqs.~(\ref{Eq:tilt-angle-EDM}) and (\ref{Eq:EDM-amplitude-2}) were used to derive Eqs.~(\ref{Eq:magnetization-EDM-2}) and (\ref{Eq:magnetization-EDM-3}), respectively. The thermal polarization of $^{207}$Pb nuclei at a temperature of $T \approx 4~{\rm K}$ and $B_0 \approx 10~{\rm T}$ is:
\begin{align}
P = \tanh\prn{ \frac{\mu_n B_0}{k_B T} } \approx \frac{\mu_n B_0}{k_B T} \approx 3 \times 10^{-4}~,
\label{Eq:thermal-polarization-Pb}
\end{align}
where $\mu_n = \hbar \gamma_n \mu_N / 2$ and $\gamma_n \approx 0.58$ for $^{207}$Pb. The spin density $n \approx 10^{22}~{\rm cm^{-3}}$ and we assume $\epsilon_S \approx 10^{-2}$ is the Schiff suppression factor for $^{207}$Pb in PMN-PT, similar to $\rm{ PbTiO_3 }$ \cite{Muk05}.

A major limiting factor in the measurement sensitivity is the magnetometer detecting $\vec{M}(t)$ as shown in Fig.~\ref{Fig:ExptSchematic}. For frequencies $10~{\rm Hz} \lesssim \omega_a/(2\pi) \lesssim 10^6~{\rm Hz}$, SQUID magnetometers offer the best sensitivity ($\approx 10^{-15}~{\rm T/\sqrt{Hz}}$), while for frequencies $\gtrsim 10^6~{\rm Hz}$ an inductive pick-up coil connected to an RF amplifier offers superior sensitivity, although potentially advantageous alternative atomic magnetometry schemes were considered in Ref.~\cite{Wan17}. To reduce noise and systematic errors, multiple samples with different orientations of their internal electric fields $\vec{E}^*$ will be used. This allows rejection of many types of common-mode noise that all samples would share, such as vibrations or uniform magnetic field noise. Samples with opposite orientations should exhibit oscillating axion-induced transverse magnetizations $180^\circ$ out-of-phase.

An important effect to consider in CASPEr Electric is chemical shift anisotropy (CSA). Chemical shift refers to the shift of the NMR frequency of a nucleus from a reference NMR frequency for that particular nucleus (measured under standard conditions). This shift is primarily produced by the local distribution of electron currents near the nucleus in question, and therefore varies as the chemical structure/environment varies. There is both isotropic chemical shift, an overall offset of the NMR frequency independent of the orientation of $\vec{B}_0$, as well as anisotropic chemical shift that depends on the relative orientation of $\vec{B}_0$ with respect to the local environment (crystal axes). For a single-crystal sample, the CSA can be fully characterized and, in principle, relatively narrow NMR lines (widths $\sim 1/T_2$, where $T_2$ is the transverse spin relaxation time) can be obtained by various NMR techniques.

The orange-shaded region in Fig.~\ref{Fig:CASPEr-Electric-sensitivity} shows the parameter space to which CASPEr Electric Phase I will be sensitive based on the above estimates and an overall integration time of $10^6~{\rm s}$. Phases II and III of CASPEr Electric rely on improving several important experimental parameters:
\begin{enumerate}

\item{increasing the degree of nuclear polarization $P$ by using optical pumping and other hyperpolarization techniques;}

\item{increasing the nuclear spin relaxation time $T_2$ by using decoupling protocols;}

\item{implementing resonant circuit detection schemes;}

\item{increasing the sample size;}

\item{longer integration time.}

\end{enumerate}

The risk factors for CASPEr Electric Phases II and III are mainly technical in nature, representing uncertainties on how well the sample material can be fabricated free of paramagnetic impurities and how effectively vibrations can be controlled. The main scientific uncertainty is the achievable degree of nuclear spin hyperpolarization. The ultimate sensitivity limit is given by the nuclear spin noise of the sample (Fig.~\ref{Fig:CASPEr-Electric-sensitivity}), which has been experimentally observed in the past using similar tools \cite{Sle85} and appears to be a feasible long-term sensitivity goal.

\section{CASPEr Wind}
\label{sec:CASPEr-Wind}

In CASPEr Wind, no electric field is required and so different possibilities for the choice of the sample are opened. Otherwise, the procedure is similar to that for CASPEr Electric: the sample is placed within a magnetic field $\vec{B}_0$; as $\vec{B}_0$ is scanned, the corresponding Larmor frequency changes; and if $\Omega_L$ is tuned to resonance with the axion oscillation frequency, an oscillating magnetization $\vec{M}(t)$ will build up in the sample. The oscillating magnetization is detected with a pick-up loop connected to a SQUID or RF amplifier.

The sample of choice for CASPEr Wind is liquid $^{129}$Xe, a high-density sample that can be hyperpolarized through spin-exchange with optically pumped Rb \cite{Rom01}. Because $^{129}$Xe is in the liquid phase, the environment is isotropic on average which removes CSA. The transverse spin relaxation times $T_2$ for liquid $^{129}$Xe can be on the order of 1000~s and over essentially the entire range of axion masses to be investigated by CASPEr, the factor limiting the integration time will be $\tau_a$.

\begin{figure*}
\includegraphics[scale=.45]{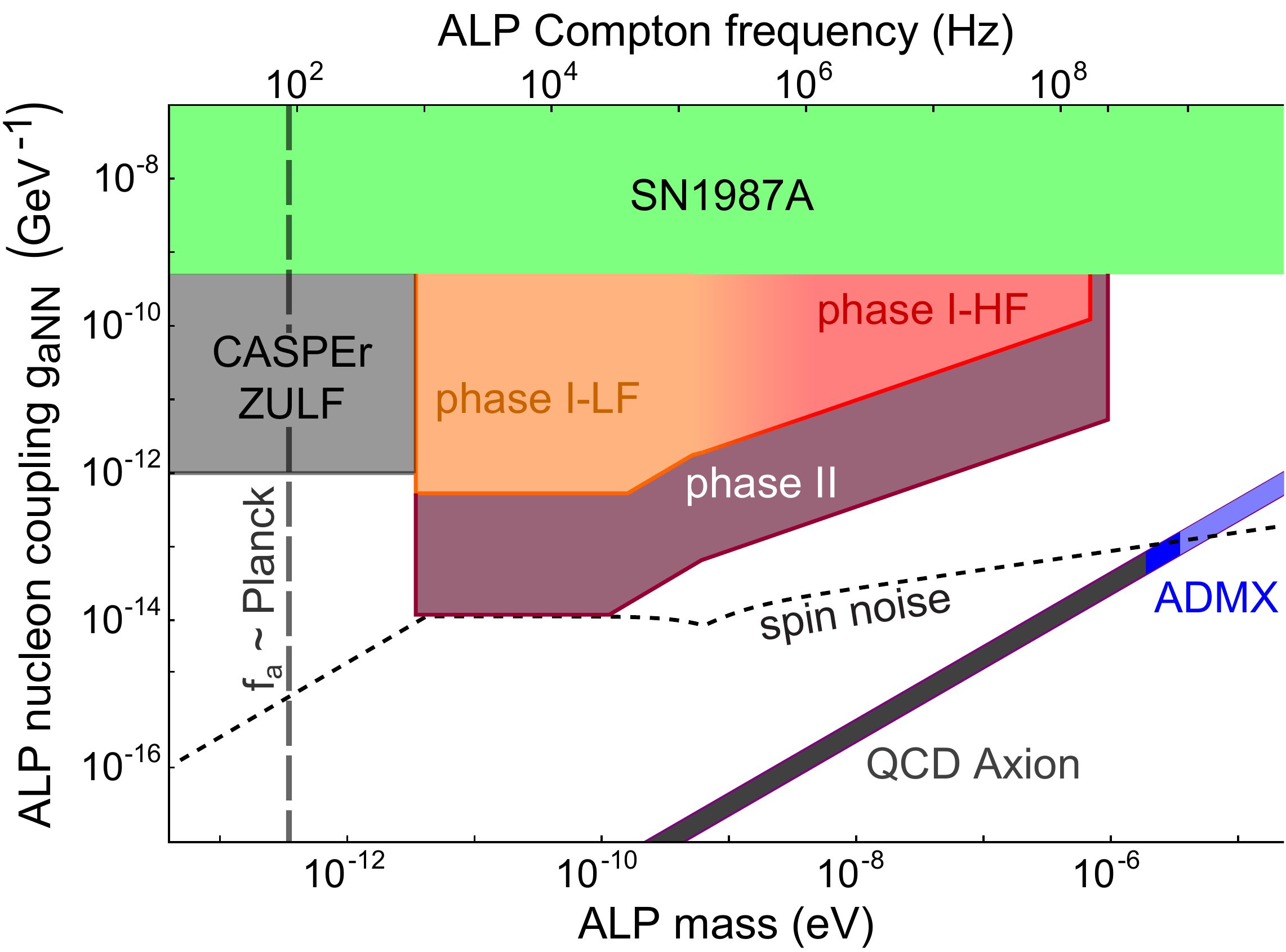}
\caption{Experimental reach of CASPEr Wind. Red, orange, and gray regions show sensitivity projections for High Field (HF), Low Field (LF), and ZULF experiments.}
\label{Fig:CASPEr-Wind-sensitivity}
\end{figure*}

As discussed in Sec.~\ref{sec:CASPEr-Electric}, different detectors are optimal for different frequency (and therefore, magnetic field) ranges. Above $\approx 1~{\rm MHz}$ ($B_0 \approx 0.1~{\rm T}$ for $^{129}$Xe), standard inductive detection using an LC circuit gives optimal signal-to-noise. However, below $\approx 1~{\rm MHz}$, SQUID magnetometers perform better. At near-zero fields corresponding to $\omega_a/(2\pi) \lesssim 10~{\rm Hz}$, other experimental strategies become viable. Thus the CASPEr Wind experiment is being realized with three distinct setups: CASPEr Wind High Field (magnetic fields $0.1~{\rm T} \lesssim B_0 \lesssim 14~{\rm T}$), CASPEr Wind Low Field ($10^{-4}~{\rm T} \lesssim B_0 \lesssim 10^{-1}~{\rm T}$), and CASPEr Wind ZULF (zero-to-ultralow field) that probes $B_0 \lesssim 10^{-4}~{\rm T}$.

\subsection{CASPEr Wind: High and Low Field}
\label{sec:CASPEr-HF-LF}

For the CASPEr Wind High and Low Field experiments the preparation of the hyperpolarized liquid Xe sample is identical (and in fact both experiments use the same Xe polarizer). To prepare the spin-polarized liquid Xe sample, first Xe gas is mixed with other gases (${\rm N_2}$, He, and Rb) and then hyperpolarized via spin-exchange optical pumping (SEOP, see Refs.~\cite{Wal97,App98}). The hyperpolarized Xe is condensed into solid form inside a region cooled by a liquid-nitrogen bath in the presence of a leading magnetic field. He and ${\rm N_2}$ used in SEOP are then vented out. Subsequently the frozen Xe is sublimated to become a gas and the valve to the experimental cell is opened and a piston compresses the Xe into liquid form. The liquid Xe is then flowed into the experimental cell. After the Xe polarization decays, the Xe is pumped out and recycled. This procedure has been demonstrated to achieve near unity polarization and spin densities on the order of $10^{22}$ per ${\rm cm^3}$ \cite{Rom01,Rus06}. The design of the experiment should allow nearly continuous cycling between the Xe polarizer and the experimental cell, enabling a high duty cycle for the measurements.

For CASPEr Wind High Field, a tunable magnetic field of up to $\approx {\rm 14~T}$ will be applied to the liquid Xe with a cryogen-free superconducting magnet, thus enabling access to axion frequencies of up to $\approx 60~{\rm MHz}$. One of the technical challenges to be addressed is maintaining the required magnetic-field homogeneity over the course of a scan. This challenge can be addressed by dynamic shimming techniques aided by in-situ-magnetic-field gradient measurements with additional magnetometers. Another possible approach is to take advantage of magnetic field gradients to multiplex the measurement and operate the experiment in analogy with a magnetic resonance imaging (MRI) measurement where different regions of the sample access different $\omega_a$.

For CASPEr Wind Low Field, the magnet requirements are less challenging, thus allowing for more rapid development of the experiment. CASPEr Wind Low Field will employ a sweepable superconducting magnet assembly inside a superconducting magnetic shield.

The projected sensitivities of Phase I CASPEr Wind High Field and Low Field experiments are shown in Fig.~\ref{Fig:CASPEr-Wind-sensitivity}. The sensitivity of CASPEr Wind Phase II is improved by significantly increasing the sample size: for Phase I the sample volume is $\approx 1~{\rm cm^3}$.

\subsection{CASPEr Wind: Zero-to-Ultralow Field (ZULF)}
\label{sec:CASPEr-ZULF}

An intriguing new possibility for a CASPEr Wind experiment at ultralow magnetic fields has recently emerged based on ZULF NMR (see Refs.~\cite{Led13,Bla17} for reviews). The basic idea of ZULF NMR is that the usual requirement of large magnetic fields for nuclear polarization, information encoding, and read-out is alleviated by the use of hyperpolarization techniques, encoding based on intrinsic spin-spin interactions, and detection methods not based on Faraday induction (e.g., SQUIDs and atomic magnetometers). ${\rm ^{15}N,^{13}C_2}$-acetonitrile (${\rm ^{13}CH_3^{13}C^{15}N}$) appears to be a suitable candidate for a CASPEr Wind sample, since non-hydrogenative parahydrogen-induced polarization (NH-PHIP) has been shown \cite{The12} to produce nuclear spin polarizations on the order of $10\%$. Additional benefits of this experimental approach is that it is relatively low-cost and non-destructive (the isotopically enriched acetonitrile can be reused indefinitely), and it enables parallel observation of three different frequencies simultaneously [the gyromagnetic ratios for the probed nuclei are $\gamma_n\prn{\rm ^1H} = 42.57~{\rm MHz/T}$, $\gamma_n\prn{\rm ^{13}C} = 10.71~{\rm MHz/T}$, and $\gamma_n\prn{\rm ^{15}N} = -4.32~{\rm MHz/T}$], which reduces the scanning time and may be advantageous for understanding and mitigating systematic effects.

Low-field NH-PHIP experiments take advantage of matching conditions that occur in ${\rm \mu T}$ magnetic fields between hydride $^1$H spins and other nuclear spins in transient iridium ``polarization-transfer'' catalysts. Essentially, parahydrogen is bubbled into a solution containing an iridium catalyst and the analyte molecule. The Ir forms a complex wherein the parahydrogen, which is in a singlet state, transfers its spin-order to other nuclear spins via electron-mediated indirect dipole-dipole coupling (J-coupling). Because the binding of parahydrogen and the analyte ligands is reversible (residence times on the order of a ms), many parahydrogen molecules are brought into contact with many analyte molecules over the bubbling timescale, allowing for a buildup of substantial non-thermal spin populations over several seconds. Which spin states are populated is dependent on the specific spin-topology of the parahydrogen-iridium-analyte complex, but it has been consistently shown that polarization enhancements of 4-5 orders of magnitude can be achieved \cite{The12}.

If a version of this experiment is performed at zero or ultralow field (the ZULF regime, $B \lesssim 1~{\rm \mu T}$), two-spin order having no net magnetization will be produced. This spin order may be converted to magnetization via sequences of magnetic field pulses. Alternatively, it may be possible to search for oscillating magnetization produced by selective population of nuclear spin states by the oscillating axion field. Because the singlet spin order has no net magnetization, it is immune to decoherence due to dipole-dipole interactions. Such long-lived nuclear spin states (demonstrated in other systems, such as $^{13}$C-formic acid at zero field \cite{Emo14}) may allow for longer integration times and longer experimental runs between repolarization.

Another intriguing aspect of the CASPEr ZULF experiment is that one can use a nonresonant measurement consisting of searching for sidebands induced by modulation of $\Omega_L$ by the axion wind interaction, removing the need to scan for the resonance. This is discussed in detail in Ref.~\cite{Gar17}. The potential sensitivity of the CASPEr ZULF experiments is shown in Fig.~\ref{Fig:CASPEr-Wind-sensitivity}.

\section{Conclusion}
\label{sec:conclusion}

The CASPEr program involves a multi-pronged experimental strategy employing NMR techniques to search for the coupling of axion dark matter to nuclear spins. First generation experiments have the potential to probe a vast range of unexplored ALP parameter space. Future generations of CASPEr experiments should achieve sufficient sensitivity to search for the QCD axion.

\section*{Acknowledgments}

The authors are grateful for the generous support of the Simons and Heising-Simons Foundations, the European Research Council, DFG Koselleck, and the National Science Foundation.


\begin{thebibliography}{99}

\bibitem{Pec77a}
Peccei, R., and Quinn, H.: Constraints imposed by $CP$ conservation in the presence of pseudoparticles. Phys. Rev. D {\textbf{16}}, 1791 (1977).

\bibitem{Pec77b}
Peccei, R., and Quinn, H.: $CP$ conservation in the presence of pseudoparticles. Phys. Rev. Lett. {\textbf{38}}, 1440 (1977).

\bibitem{Bak06}
Baker, C. A., Doyle, D. D., Geltenbort, P., Green, K., van der Grinten, M. G. D., Harris, P. G., Iaydjiev, P., Ivanov, S. N., May, D. J. R., Pendlebury, J. M., Richardson, J. D., Shiers, D., and Smith, K. F.: Improved experimental limit on the electric dipole moment of the neutron. Phys. Rev. Lett. {\textbf{97}}, 131801 (2006).

\bibitem{Wei78}
Weinberg, S. A.: New Light Boson? Phys. Rev. Lett. {\textbf{40}}, 223 (1978).

\bibitem{Wil78}
Wilczek, F.: Problem of Strong $P$ and $T$ Invariance in the Presence of Instantons. Phys. Rev. Lett. {\textbf{40}}, 279 (1978).

\bibitem{Kim79}
Kim, J. E.: Weak-interaction singlet and strong $CP$ invariance. Phys. Rev. Lett. {\textbf{43}}, 103 (1979).

\bibitem{Shi80}
Shifman, M. A., Vainshtein, A. I., and Zakharov, V. I.: Can confinement ensure natural $CP$ invariance of strong interactions? Nucl. Phys. B {\textbf{166}}, 493 (1980).

\bibitem{Din81}
Dine, M., Fischler, W., and Srednicki, M.: A simple solution to the strong $CP$ problem with a harmless axion. Phys. Lett. {\textbf{104B}}, 199 (1981).

\bibitem{Zhi80}
Zhitnitsky, A. R.: Weinberg's model of $CP$ violation and $T$-odd correlations in weak decays. Sov. J. Nucl. Phys. {\textbf{31}}, 529 (1980).

\bibitem{Svr06}
Svrcek, P., and Witten, E.: Axions in string theory. J. High Energy Phys. {\textbf{06}}, 051 (2006).

\bibitem{Arv10}
Arvanitaki, A., Dimopoulos, S., Dubovsky, S., Kaloper, N., and March-Russell, J.: String axiverse. Phys. Rev. D {\textbf{81}}, 123530 (2010).

\bibitem{Din83}
Dine, M. and Fischler, W.: The not-so-harmless axion. Phys. Lett. {\textbf{120B}}, 137 (1983).

\bibitem{Pre83}
Preskill, J., Wise, M. B., and Wilczek, F.: Cosmology of the invisible axion. Phys. Lett. {\textbf{120B}}, 127 (1983).

\bibitem{Ber98}
Bergstr\"om, L., Ullio, P., and Buckley, J. H.: Observability of $\gamma$ rays from dark matter neutralino annihilations in the Milky Way halo. Astroparticle Physics {\textbf{9}}, 137 (1998).

\bibitem{Jun96}
Jungman, G., Kamionkowski, M., and Griest, K. Supersymmetric dark matter. Phys. Rep. {\textbf{267}}, 195 (1996).

\bibitem{Sof01}
Sofue, Y., and Rubin, V.: Rotation curves of spiral galaxies. Annual Review of Astronomy and Astrophysics {\textbf{39}}, 137 (2001).

\bibitem{Gra13}
Graham, P. W., and Rajendran, S.: New observables for direct detection of axion dark matter. Phys. Rev. D {\textbf{88}}, 035023 (2013).

\bibitem{Sik83}
Sikivie, P.: Experimental tests of the ``invisible'' axion. Phys. Rev. Lett. {\textbf{51}}, 1415 (1983).

\bibitem{Sik85}
Sikivie, P.: Detection rates for ``invisible''-axion searches. Phys. Rev. D {\textbf{32}}, 2988 (1985).

\bibitem{Vys78}
Vysotsky, M. I., Zeldovich, Ya. B., Khlopov, M. Yu., and Chechetkin, V. M.:
Some astrophysical limitations on the axion mass. Pis' ma Zh. Eksp. Teor. Fiz. {\textbf{27}}, 533 (1978); English translation: JETP Lett. {\textbf{27}}, 502 (1978).

\bibitem{Ber91}
Berezhiani, Z. G. and Khlopov, M. Yu.: Cosmology of spontaneously broken gauge family symmetry with axion solution of strong CP-problem. Z. Phys. C {\textbf{49}}, 73 (1991).

\bibitem{Sak96}
Sakharov, A. S., Sokoloff, D. D., and Khlopov, M. Yu.: Large scale modulation of the distribution of coherent oscillations of a primordial axion field in the Universe. Yadernaya Fizika {\textbf{59}}, 1050 (1996); English translation: Phys. Atom. Nucl. {\textbf{59}}, 1005 (1996).

\bibitem{Khl99}
Khlopov, M. Yu., Sakharov, A. S. and Sokoloff, D. D.: The nonlinear modulation of the density distribution in standard axionic CDM and its cosmological impact. Nucl.Phys. B (Proc. Suppl.) {\textbf{72}}, 105 (1999).

\bibitem{Raf88}
Raffelt, G., and Seckel, D.: Bounds on exotic-particle interactions from SN1987A. Phys. Rev. Lett. {\textbf{60}}, 1793 (1988).

\bibitem{Raf95}
Raffelt, G. and Weiss, A.: Red giant bound on the axion-electron coupling reexamined. Phys. Rev. D {\textbf{51}}, 1495 (1995).

\bibitem{Raf12}
Raffelt, G.: Limits on a CP-violating scalar axion-nucleon interaction. Phys. Rev. D {\textbf{86}}, 015001 (2012).

\bibitem{Blu16}
K. Blum and D. Kushnir. Neutrino signal of collapse-induced thermonuclear supernovae: The case for prompt black hole formation in SN 1987A. The Astrophysical Journal, {\textbf{828}}, 31 (2016).

\bibitem{Cha18}
Chang, J. H.,  Essig, R., and McDermott, S. D.: Supernova 1987a constraints on sub-GeV dark sectors, millicharged particles, the QCD axion, and an axion-like particle. J. High Energ. Phys. {\textbf{2018}}, 51 (2018).

\bibitem{Saf18}
Safranova, M. S., Budker, D., DeMille, D., Jackson Kimball, D. F., Derevianko, A., and Clark, C. W.: Search for new physics with atoms and molecules. Rev. Mod. Phys. {\textbf{90}}, 025008 (2018).

\bibitem{Gra15}
Graham, P. W., Irastorza, I. G., Lamoreaux, S. K., Lindner, A., and van Bibber, K. A.: Experimental searches for the axion and axion-like particles. Annu. Rev. Nucl. Part. Sci. {\textbf{65}}, 485 (2015).

\bibitem{Raf99}
Raffelt, G. G.: Particle physics from stars. Annu. Rev. Nucl. Part. Sci. {\textbf{49}}, 163 (1999).

\bibitem{Asz01}
Asztalos, S., Daw, E., Peng, H., Rosenberg, L. J., Hagmann, C., Kinion, D., Stoeffl, W., van Bibber, K., Sikivie, P.,  Sullivan, N. S., Tanner, D. B., Nezrick, F., Turner, M. S., Moltz, D. M., Powell, J., Andr\'e, M.-O., Clarke, J., M\"uck, M., and Bradley, R. F.: Large-scale microwave cavity search for dark-matter axions. Phys. Rev. D {\textbf{64}}, 092003 (2001).

\bibitem{Asz10}
Asztalos, S. J., Carosi, G., Hagmann, C., Kinion, D., van Bibber, K., Hotz, M., Rosenberg, L. J., Rybka, G., Hoskins, J., Hwang, J., Sikivie, P., Tanner, D. B., Bradley, R., and Clarke, J.: Squid-based microwave cavity search for dark-matter axions. Phys. Rev. Lett. {\textbf{104}}, 041301 (2010).

\bibitem{Bru17}
Brubaker, B. M., Zhong, L., Gurevich, Y. V., Cahn, S. B., Lamoreaux, S. K., Simanovskaia, M., Root, J. R., Lewis, S. M., Al Kenany, S., Backes, K. M., Urdinaran, I., Rapidis, N. M., Shokair, T. M., van Bibber, K. A., Palken, D. A., Malnou, M., Kindel, W. F., Anil, M. A., Lehnert, K. W., and Carosi, G.: First Results from a Microwave Cavity Axion Search at 24~$\mu$eV. Phys. Rev. Lett. {\textbf{118}}, 061302 (2017).

\bibitem{You16}
Youn, S.: Axion research at CAPP/IBS. International Journal of Modern Physics: Conference Series {\textbf{43}}, 1660193 (2016).

\bibitem{Bud14}
Budker, D., Graham, P. W., Ledbetter, M., Rajendran, S., and Sushkov, A. O.: Proposal for a Cosmic Axion Spin Precession Experiment (CASPEr). Phys. Rev. X {\textbf{4}}, 021030 (2014).

\bibitem{Bud08}
Budker, D., Kimball, D. F., and DeMille, D. P.: Atomic physics: an exploration through problems and solutions. Oxford University Press, USA (2008).

\bibitem{Rom01}
Romalis, M. V. and Ledbetter, M. P.: Transverse Spin Relaxation in Liquid Xe-129 in the Presence of Large Dipolar Fields. Phys. Rev. Lett. {\textbf{87}}, 067601 (2001).

\bibitem{Gar17}
Garcon, A., Aybas, D., Blanchard, J. W., Centers, G., Figueroa, N. L., Graham, P., Jackson Kimball, D. F., Rajendran, S., Sendra, M. G., Sushkov, A. O., Trahms, L., Wang, T. , Wickenbrock, A., Wu, T., and Budker, D.: The Cosmic Axion Spin Precession Experiment (CASPEr): a dark-matter search with nuclear magnetic resonance. Quantum Science and Technology {\textbf{3}}, 014008 (2018).

\bibitem{Lin88}
Linde, A. D.: Inflation and axion cosmology. Phys. Lett. B {\textbf{201}}, 437 (1988).

\bibitem{Lin07}
Linde, A. D.: Towards a gauge invariant volume-weighted probability measure for eternal inflation. J. Cosmol. Astropart. Phys. {\textbf{06}}, 017 (2007).

\bibitem{McD13}
McDonald, J.: Explaining the dark energy, baryon and dark matter coincidence via domain-dependent random densities. J. Cosmol. Astropart. Phys. \textbf{2013(05)}, 019 (2013).

\bibitem{Gra15relaxion}
Graham, P. W., Kaplan, D .E., and Rajendran, S.: Cosmological relaxation of the electroweak scale. Phys. Rev. Lett. {\textbf{115}}, 221801 (2015).

\bibitem{Sha68}
Shapiro, F. L.: Electric Dipole Moments of Elementary Particles. Sov. Phys. Usp. {\textbf{11}}, 345 (1968).

\bibitem{Leg78}
Leggett, A. J.: Macroscopic effect of $P$-and $T$-nonconserving interactions in ferroelectrics: a possible experiment? Phys. Rev. Lett. {\textbf{41}}, 586 (1978).

\bibitem{Muk05}
Mukhamedjanov, T. N. and Sushkov, O. P.: A Suggested Search for Pb-207 Nuclear Schiff Moment in PbTiO$_3$ Ferroelectric. Phys. Rev. A {\textbf{72}}, 034501 (2005).

\bibitem{Bud06}
Budker, D., Lamoreaux, S. K., Sushkov, A. O., and Sushkov, O. P.: Sensitivity of Condensed Matter $P$- and $T$-Violation Experiments, Phys. Rev. A {\textbf{73}}, 022107 (2006).

\bibitem{Sus10}
Sushkov, A. O., Eckel, S. and Lamoreaux, S. K.: Prospects for an Electron Electric-Dipole-Moment Search with Ferroelectric (Eu,Ba)TiO$_3$ Ceramics. Phys. Rev. A {\textbf{81}}, 022104 (2010).

\bibitem{Rus10}
Rushchanskii, K. Z., Kamba, S., Goian, V., Van$\check{\rm e}$k, P., Savinov, M., Prokle$\check{\rm s}$ka, J., Nuzhnyy, D., Kn\'i$\check{\rm z}$ek, K., Laufek, F., Eckel, S., Lamoreaux, S. K., Sushkov, A. O., Le$\check{\rm z}$ai\'c, M., and Spaldin, N. A.: A Multiferroic Material to Search for the Permanent Electric Dipole Moment of the Electron, Nat. Mater. {\textbf{9}}, 649 (2010).

\bibitem{Eck12}
Eckel, S., Sushkov, A. O. and Lamoreaux, S. K.: Limit on the Electron Electric Dipole Moment Using Paramagnetic Ferroelectric Eu$_{0.5}$Ba$_{0.5}$TiO$_3$. Phys. Rev. Lett. {\textbf{109}}, 193003 (2012).

\bibitem{Bar14}
Baron, J., Campbell, W. C., DeMille, D., Doyle, J. M., Gabrielse, G., Gurevich, Y. V., Hess, P. W., Hutzler, N. R., Kirilov, E., Kozyryev, I., O'Leary, B. R., Panda, C. D., Parsons, M. F., Petrik, E. S., Spaun, B., Vutha, A. C., and West, A. D.: Order of magnitude smaller limit on the electric dipole moment of the electron. Science {\textbf{343}}, 269 (2014).

\bibitem{Sch63}
Schiff, L.I.: Measurability of nuclear electric dipole moments. Phys. Rev. {\textbf{132}}, 2194 (1963).

\bibitem{San65}
Sandars, P.G.H.: The electric dipole moment of an atom. Phys. Lett. {\textbf{14}}, 194 (1965).

\bibitem{Mol68}
Mollenauer, L. F., Grant, W. B. and Jeffries, C. D.: Achievement of Significant Nuclear Polarizations in Solids by Optical Pumping. Phys. Rev. Lett. {\textbf{20}}, 488 (1968).

\bibitem{Hay08}
Hayes, S. E., Mui, S., and Ramaswamy, K.: Optically pumped nuclear magnetic resonance of semiconductors. J. Chem. Phys. {\textbf{128}}, 052203 (2008).

\bibitem{Bou08}
Bouchard, L. S., Sushkov, A. O., Budker, D., Ford, J. J., and Lipton, A. S.: Nuclear-spin relaxation of Pb-207 in ferroelectric powders. Phys. Rev. A {\textbf{77}}, 022102 (2008).

\bibitem{Wan17}
Wang, T., Jackson Kimball, D. F., Sushkov, A. O., Aybas, D., Blanchard, J. W., Centers, G., Kelley, S. R., Fang, J. and Budker, D.: Application of Spin-Exchange Relaxation-Free Magnetometry to the Cosmic Axion Spin Precession Experiment. Physics of the Dark Universe {\textbf{19}}, 27 (2018).

\bibitem{Sle85}
Sleator, T., Hahn, E., Hilbert, C., and Clarke, J.: Nuclear-spin noise. Phys. Rev. Lett. {\textbf{55}}, 1742 (1985).

\bibitem{Wal97}
Walker, T. G. and Happer, W.: Spin-exchange optical pumping of noble-gas nuclei. Rev. Mod. Phys. {\textbf{69}}, 629 (1997).

\bibitem{App98}
Appelt, S., Baranga, A. B .A., Erickson, C. J., Romalis, M. V., Young, A. R., and Happer, W.: Theory of spin-exchange optical pumping of He-3 and Xe-129. Phys. Rev. A {\textbf{58}}, 1412 (1998).

\bibitem{Rus06}
Ruset, I. C., Ketel, S., and Hersman, F. W.: Optical pumping system design for large production of hyperpolarized Xe-129. Phys. Rev. Lett. {\textbf{96}}, 053002 (2006).

\bibitem{Led13}
Ledbetter, M. P. and Budker, D.: Zero-field nuclear magnetic resonance. Phys. Today, {\textbf{66}}, 44 (2013).

\bibitem{Bla17}
Blanchard, J. W. and Budker, D.: Zero-to-Ultralow-Field NMR. eMagRes (2016).

\bibitem{The12}
Theis, T., Ledbetter, M. P., Kervern, G., Blanchard, J. W., Ganssle, P. J., Butler, M. C., Shin, H. D., Budker, D., and Pines, A.: Zero-field NMR enhanced by parahydrogen in reversible exchange. J. Am. Chem. Soc. {\textbf{134}}, 3987 (2012).

\bibitem{Emo14}
Emondts, M., Ledbetter, M. P., Pustelny, S., Theis, T., Patton, B., Blanchard, J. W., Butler, M. C., Budker, D., and Pines, A.: Long-Lived Heteronuclear Spin-Singlet States in Liquids at a Zero Magnetic field. Phys. Rev. Lett. {\textbf{112}}, 077601 (2014).



\end{thebibliography}
\end{document}